\documentclass[10pt]{article}

\usepackage{amsfonts,a4wide}

\usepackage{amsmath,amssymb,amsthm}
\usepackage{url}
\newtheorem{Th}{Theorem}

\newtheorem{Lemma}{Lemma}
\newtheorem{Cor}{Corollary}

\theoremstyle{remark}

\newtheorem{Rem}{Remark}

\newcommand{\const}{\mbox{\rm const}}

\newcommand{\bbR}{\mathbb{R}}

\newcommand{\weg}[1]{}

\begin{document}
\title{Normal forms for pseudo-Riemannian 2-dimensional  metrics whose geodesic flows admit  integrals quadratic in momenta}
\author{
Alexey V. Bolsinov\thanks{
Department of Mathematical Sciences,
Loughborough University,  LE11 3TU
UK, A.Bolsinov@lboro.ac.uk},
Vladimir S. Matveev\thanks{Institute of Mathematics, FSU Jena, 07737 Jena Germany,  matveev@minet.uni-jena.de},
Giuseppe Pucacco\thanks{Dipartimento di Fisica, Universit\`a di Roma ``Tor Vergata", 00133 Rome Italy, pucacco@roma2.infn.it}}
\date{}
\maketitle

\begin{abstract}
We discuss pseudo-Riemannian metrics on 2-dimensional manifolds such
that the geodesic flow admits a nontrivial integral quadratic in
velocities.  We construct (Theorem \ref{main}) local normal forms of
such metrics. We show that these metrics have  certain useful
properties
 similar to those  of Riemannian Liouville   metrics, namely:  \\
$\bullet$ they admit geodesically equivalent metrics
 (Theorem \ref{thm3});\\
$\bullet$
 one can use them to construct a large family of natural  systems admitting  integrals quadratic in momenta (Theorem \ref{potential}); \\
$\bullet$ the integrability of such systems can be generalized to the quantum setting (Theorem \ref{quantum});    \\
$\bullet$ these natural  systems are integrable by quadratures (Section \ref{har}).

\end{abstract}

\section{Introduction}

Consider a  pseudo-Riemannian metric $g=(g_{ij})$ on  a surface
$M^2$.  A function $F:{T^*}M\to \mathbb{R}$ is called {\bf an
integral} of the geodesic flow of $g$, if $\{ H, F\}=0$, where $H:=
\tfrac{1}{2} g^{ij} p_ip_j:T^*M\to \mathbb{R}$ is the kinetic energy
corresponding to the metric. Geometrically, this condition means
that the function is constant on the orbits of the Hamiltonian
system with the Hamiltonian $H$.  We say the  integral $F$ is {\bf
quadratic in momenta} if, in every local coordinate system $(x,y)$
on $M^2$,  it has the form
\begin{equation} \label{integral}
a(x,y)p_x^2+ b(x,y)p_xp_y+ c(x,y)p_y^2,
\end{equation}
with $(x,y,p_x, p_y)$ canonical coordinates on $T^*M^2$.
Geometrically, formula \eqref{integral} means that the restriction
of the integral to every cotangent space $T^*_pM^2\equiv
\mathbb{R}^2$  is a homogeneous quadratic function. Of course, $H$
itself  is  an integral quadratic  in momenta for $g$. We will say
that the integral $F$ is {\bf nontrivial}, if $F\ne \textrm{const}
\cdot H$ for all $\textrm{const} \in \mathbb{R}$.

The main result of this paper is  Theorem~\ref{main} below, which
gives us a list of local normal forms of  metrics  of signature
$(+,-)$ whose geodesic flows  admit a nontrivial integral quadratic
in momenta. For the Riemannian case (and, therefore, for the
signature $(-,-)$)  such  metrics are the well-known Liouville
metrics.

\begin{Th} \label{main}   Suppose the metric $g$ of  signature       $(+,-)$  on ${M}^2$ admits a nontrivial integral quadratic in momenta. Then, in a neighbourhood of almost every point there exist coordinates  $x,y$ such that  the metric and the integral are as in the following table:
\begin{center}\begin{tabular}{|c||c|c|c|}\hline & \textrm{Liouville case} & \textrm{Complex-Liouville case} & \textrm{Jordan-block case}\\ \hline \hline
$g$ & $(X(x)-Y(y))(dx^2 -dy^2)$ &  $\Im(h)dxdy$ & $\left(1+{x} Y'(y)\right)dxdy $
\\ \hline  $F$&$\tfrac{X(x)p_y^2 -Y(y)p_{x}^2}{X(x)-Y(y)} $&
 $p_x^2 - p_y^2   + 2\tfrac{\Re(h)}{\Im(h)}p_xp_y $ & $p_x^2 -2  \frac{Y(y)}{1+ {x}{} Y'(y) }p_xp_y$  \\ \hline
\end{tabular}
\end{center}
where  $\Re (h)$ and $\Im (h)$ are the real and imaginary parts of a
holomorphic function $h$ of the variable $z:= x+ i\cdot y$.
\end{Th}

Given a metric and  the quadratic  integral, it is easy to understand  what case they belong to. Indeed, for the integral \eqref{integral}
 the matrix
 $$
 F^{ij}= \begin{pmatrix} a & \tfrac{b}{2} \\ \tfrac{b}{2} & c \end{pmatrix}
 $$
  can be viewed as  a $(2,0)$-tensor: if we change the coordinate system and rewrite the  function $F$ in the new coordinates, the matrix changes according to the tensor rule. Then,
  \begin{equation} \label{matrix}
  G_j^{i}:= \sum_{\alpha} g_{j\alpha} F^{i\alpha}
  \end{equation}
  is a $(1,1)$-tensor.  By direct   calculation we see that $G_j^i$ has two different real eigenvalues in the first case, two complex-conjugate eigenvalues in the second case and is (conjugate to) a Jordan-block in the third case. This also explains our choice of the names for the normal forms of the metrics. Indeed, in  the Riemannian case, the tensor
  \eqref{matrix}  always has two real eigenvalues. In particular, the normal  form of the Riemannian metric admitting an integral quadratic in momenta, which is traditionally called Liouville form (or Liouville metric),  is very similar  to the metric of our ``Liouville" case.   One can view our ``Complex-Liouville" case as the complexification of the standard Liouville metric: if  in the expression
  $$(X(x)- Y(y)) (dx^2 + dy^2)$$
  we replace $X$ by (a holomorphic function) $h(z)$,  $Y$ by $\overline{h(z)}$, $dx$ by $dz$,  and $dy$ by $i d\bar z$, we obtain the Complex-Liouville metric up to the factor $8i$.    The Jordan-block case has no direct analog in the Riemannian setting.

 \begin{Rem} The corresponding natural Hamiltonian problem on the hyperbolic plane has recently been treated in \cite{PR} following an approach used by Rosquist and Uggla \cite{ru:kt}. Systems with indefinite signature have been investigated before in the classical works by Kalnins and Miller on separation of variables \cite{kalnins,miller}, see also \cite{chanu,rastelli,smirnov}.  Other possible approaches are based on Killing tensor theory \cite{Benenti2}, r-matrix theory \cite{KS} and algebraic methods \cite{das}. For the corresponding quantum case we refer to \cite{harnad} and references therein.
 
 \end{Rem}

 \begin{Rem} A part, if not all credits for the results of the present paper should be given  to Darboux, see  \cite[\S\S592--594,600--608]{Darboux}. There is no doubt that Darboux was very close to  Theorem \ref{main},  to the results of Section \ref{2.2.3}, and, to a certain extent, to Theorem \ref{thm3}   of our paper,  and could get it if he would have been interested  in the pseudo-Riemannain metrics.   More precisely,

  \begin{itemize} \item In \cite[\S593]{Darboux}, Darboux gets the Riemannian  Liouville metrics.  Since he worked over complex coordinates, his formulas can be interpreted as our  Liouville  and  Complex-Liouville cases.

\item  In \cite[\S594]{Darboux}, Darboux gets (a case that {could} be interpreted as) the
Jordan-block case.

\item The formulas of  Section \ref{2.2.3}  of the present paper are similar  to that of \cite[\S594]{Darboux}.

\end{itemize}

However,    Darboux  was interested in the positive definite metrics
only. Actually, in his time it was unusual to consider indefinite
metrics, since the applications of pseudo-Riemannian metrics to
general relativity and cosmology appeared much later.      Darboux
worked over complex coordinates $x,y$  and explicitly remarked on
the transformation $x = u + iv, y = u - iv$ leading to the standard
metric of the (+,+) case, with no mention of a possible
interpretation of $x,y$ as {real} coordinates.   The only exception
is the Jordan-block case with constant function $Y$ (equations
(24,25) of \cite[\S594]{Darboux}), where one can get the surfaces of
revolution.

The results of this paper were announced in \cite{BMP}.

\end{Rem}

\section{Applications}

\subsection{Applications in geometry: normal forms for 2-dimensional  geodesically equivalent metrics}

 Two metrics $g$ and $\bar g$ on one manifold are {\bf geodesically equivalent,} if every  (unparametrized) geodesic of the first metric is a geodesic  of the second metric. Investigation of geodesically equivalent metrics  is a classical problem in differential geometry, see the surveys \cite{Aminova2,fom,mikes}  or/and the introductions to \cite{threemanifolds,hyperbolic,diffgeo}. In particular, normal forms for geodesically equivalent Riemannian 2-dimensional  metrics were already constructed by Dini \cite{Dini}.  An easy corollary of Theorem~\ref{main} is the following theorem which gives normal forms of geodesically equivalent nonproportional metrics such that one of them  has signature $(+,-)$.

\begin{Th} \label{thm3}
Let $g$, $\bar g$ be geodesically equivalent  metrics on $M^2$ such
that $g$ has signature $(+, -)$, and $\bar g\ne \textrm{const} \cdot
g$ for every $\textrm{const}\in \mathbb{R}$. Then, in
a~neighbourhood of almost every point, there exist coordinates such
that metrics are as in the following table:
\begin{center}
 \begin{tabular}{|c||c|c|c|}\hline &  \textrm{Liouville case} & \textrm{Complex-Liouville case} & \textrm{Jordan-block case}\\ \hline \hline
$g$ & $(X(x)-Y(y))(dx^2 -dy^2)$ &  $\Im(h)dxdy$ & $\left( 1+{x} Y'(y)\right)dxdy $
\\  \hline  $ \bar g$  &$ \left( \frac{1}{Y(y)}-\frac{1}{X(x)}\right) \left( \frac{dx^2}{X(x)} -  \frac{dy^2}{Y(y)} \right)$&
 \begin{minipage}{.3\textwidth}$-\left(\frac{\Im(h)}{\Im(h)^2 +\Re(h)^2}\right)^2dx^2 \\   +2\frac{\Re(h) \Im(h)}{   (\Im(h)^2 +\Re(h)^2)^2} dx dy  \\ +  \left(\frac{\Im(h)}{\Im(h)^2 +\Re(h)^2}\right)^2dy^2 $
 \end{minipage} &  \begin{minipage}{.3\textwidth}$  \frac{1+{x} Y'(y)}{Y(y)^4} \bigl(- 2Y(y) dxdy\\
    + (1+{x} Y'(y))dy^2\bigr)$\end{minipage}\\ \hline
\end{tabular}
\end{center}
where $h$ is  holomorphic function of the variable $z:= x +i\cdot y $.   \end{Th}

\begin{Rem} 
It it natural to consider the metrics from the Complex-Liouville case as the complexification of the metrics from the Liouville case: indeed, in the complex coordinates $z= x + i\cdot y$, $\bar z= x - i\cdot y$, the metrics  have the form

 \begin{equation*}  \left. \begin{array}{ccc}   ds^2_g & =  &  -\tfrac{1}{8}( \overline{h( z}) - h(z) )\left(d\bar z^2 -   dz^2\right),\\   
     ds^2_{\bar g} & =  &  -\tfrac{1}{4}\left(\frac{1}{\overline{h( z})} - \frac{1}{h(z)} \right)\left(\frac{d\bar z^2}{ \overline{h(z})} - \frac{ dz^2}{h(z)}\right).\end{array}\right. \end{equation*}
\end{Rem} 

\begin{Rem}\label{r4}  In the Jordan-block case, 
if $dY\ne 0$ (which is always the case  at almost every point,  if the restriction of  $g$ to any neighborhood   does   not admit a Killing vector field), after a  local coordinate change, the metrics  $g$ and   $\bar g$  have the form  (see also Remark \ref{14})  \begin{eqnarray*}ds_g^2 & =&  \left( \tilde Y(y) +{x}{} \right)dxdy \\ 
   ds_{\bar g}^2   & =&  -\frac{2(\tilde Y(y)+x)}{y^3}dxdy  + \frac{(\tilde Y(y)+x)^2}{y^4}dy^2.
 \end{eqnarray*}
  \end{Rem} 
\noindent {\bf Proof of Theorem \ref{main}. }   We will use  the next theorem  which
probably was already known to    Darboux~\cite[\S608]{Darboux}.
For recent proofs, see~\cite{MT,dim2,ERA,dedicata}.

 \begin{Th} \label{thm2}
Let $g$ be a metric on $M^2 $ and $h\in \Gamma(S_2M^2)$ be a symmetric nondegenerate bilinear form on $M^2$.  Consider the following metric
\begin{equation} \label{int}
\bar g = \left(\frac{\det(g)}{\det(h)}\right)^2 h
\end{equation}
on $M^2$. If $g$ and $\bar g$ are geodesically equivalent, then the function
 $$\hat h:TM\to \bbR, \ \hat h(\xi):= h(\xi,\xi)$$ is an integral for the  geodesic flow of $g$.
\end{Th}

\begin{Rem} Theorem \ref{thm2} and Corollary \ref{cor1} below  bear some resemblance with other classes of transformations between dynamical systems \cite{abenda1,abenda,Boyer,miller,blaszak,tsiganov1,tsiganov}. However, the present result is  of different nature and is deeper because, in order to construct the second system, one needs to know the quadratic integral of the first one. \end{Rem} 

Combining Theorem \ref{thm2}  with Theorem~\ref{main},
 we obtain that, in a neighbourhood of almost every point, geodesically equivalent metrics $g$ and $\bar g$ are as in the table in Theorem~\ref{thm3}  (we assume that $g$ has signature $(+,-)$ and that $\bar g \ne \textrm{const} \cdot g$).
Thus, in order to  prove Theorem~\ref{thm3}, we need to show that the metrics from the table are indeed geodesically equivalent, which can be done by direct calculations. Indeed, it is well-known, see for example \cite[\S40 of Ch. III]{Eisenhart}, that  two metrics are geodesically equivalent if and only if the difference of their Levi-C\`{\i}vita  connections has the form $\Upsilon_j\delta^{i}_{k} +\Upsilon_k\delta^{i}_{j}  $  for a one-form $\Upsilon= \left(\Upsilon_i\right)$. Direct calculation of the Levi-C\`{\i}vita connections for the metrics shows that it is indeed the case: the form $\Upsilon $ equals
$$\frac{1}{2}\left(\frac{X'(x)}{X(x)}dx + \frac{Y'(y)}{Y(y)}dy\right)$$
for the normal forms of the metrics in the Liouville case,
\weg{$\tfrac{1}{2}\left( \left(\frac{\partial } {\partial x} \frac{1}{\Re(h)^2 +\Im(h)^2}\right)dx  +  \left(\frac{\partial } {\partial y} \frac{1}{\Re(h)^2 +\Im(h)^2}\right)dy \right)$}
$${\frac {{\Im(h)} {\frac {\partial }{
\partial x{{}}}}{\Im(h)}  + \Re(h)
   {\frac {\partial }{\partial y{{}}}}{
 \Im(h)}   }{ \left(  \Im(h) \right)^{2}+ \left(  \Re(h) \right)^{2}}}dx + {\frac { \Im(h)   {\frac {\partial }{
\partial y{{}}}} \Im(h)   - \Re(h)
   {\frac {\partial }{\partial x{{}}}}{
\Im(h)}   }{ \left(  \Im(h)   \right) ^{2}+ \left(  \Re(h)   \right) ^{2}}}dy$$
for the complex Liouville case and $\frac{Y'(y)}{Y(y)}dy$ for the  Jordan-block case. \qed

\begin{Cor} \label{cor1}
Let $g$ be a metric on $M^2 $ and $h\in \Gamma(S_2M^2)$ be a symmetric nondegenerate bilinear form on $M^2$.    Then,  $g$ and the metric \eqref{int} are geodesically equivalent, if and only if  the function
 $$\hat h:TM\to \bbR, \ \hat h(\xi)= h(\xi,\xi)$$ is an integral for the  geodesic flow of $g$.
\end{Cor}

{\bf Proof.}  In  the direction ``$\Longrightarrow$" the statement coincides with Theorem~\ref{thm2}. In order to prove in ``$\Longleftarrow$" direction,  it is sufficient to check  the statement in the neighbourhood of  almost every point. Here, the metrics $g$, $\bar g$ and the integrals $\hat h$ are given by  Theorems~\ref{main},\ref{thm3}  and are related  precisely    by formula \eqref{int}. \qed

\begin{Rem} Theorem \ref{thm2} had found a recent important application in 
 the solution  of two problems explicitly stated by Sophus Lie in \cite{Lie} due to \cite{bryant,alone}.  
\end{Rem} 

\subsection{Applications in mathematical physics}

\subsubsection{ Natural systems  admitting an integral quadratic in momenta }

For a pseudo-Riemannian manifold  $(M,g)$, a {\bf natural Hamiltonian
  system}  is a  Hamitonian system with $H:T^*M\to \mathbb{R}$   of the form $H:= H_g +  U = \tfrac{1}{2} g^{ij} p_ip_j + U(x,y)$. We say that a natural Hamiltonian system is {\bf quadratically integrable}, if there exists a function $F$ of the form $F= F_g + V = F^{ij}p_ip_j + V(x,y)$ such that $\{H, F\}=0$ with $F\ne \textrm{const}_1 \cdot  H + \textrm{const}_2$ for all  $\textrm{const}_1, \textrm{const}_2\in \mathbb{R}$.

 \begin{Rem} In \cite{PR}, the natural Hamiltonian system on the Minkowski plane has been reduced to the corresponding kinetic Hamiltonian system with conformal (Jacobi) pseudo-Euclidean metric. \end{Rem}

\begin{Th}   \label{potential}
Let $g$ be a  metric of signature $(+,-)$ on $M^2$. Assume a natural
Hamiltonian system with Hamiltonian $H_g+ U$ to be quadratically
integrable with integral $F= F_g + V$. Then, in a neighbourhood of
almost every point, there exists a coordinate system such that
  the metric $g$  and the functions $F_g $, $U$, $V$   are as in the following table:

   \begin{center} \begin{tabular}{|c||c|c|c|}\hline & \textrm{Liouville case} & \textrm{Complex-Liouville case} & \textrm{Jordan-block case}\\ \hline \hline
$g$ & $(X(x)-Y(y))(dx^2 -dy^2)$ &  $\Im(h)dxdy$ & $\left( 1+{x} Y'(y)\right)dxdy $
\\ \hline  $F_g$&$\tfrac{X(x)p_y^2 -Y(y)p_{x}^2}{X(x)-Y(y)} $&
 $p_x^2 - p_y^2   + 2\tfrac{\Re(h)}{\Im(h)}p_xp_y, $ & $p_x^2 -2  \frac{Y(y)}{1+ {x}{} Y'(y) }p_xp_y$       \\ \hline
  U &  $\tfrac{1}{2}\frac{\hat X(x) -\hat Y(y)}{X(x)- Y(y)} $& $ {}\frac{ \Im(h_1) }{\Im(h)} $  &  $\frac{ x Y_1'(y) +  Y_2(y)}{1+ xY'(y)}$  \\
  \hline V  & $\frac{ \hat Y(y)   X(x) - \hat X(x) Y(y)}{X(x)-Y(y)} $&
  $  \Re(h)\frac{ \Im(h_1) }{\Im(h)}- \Re(h_1) $ &$ -Y \frac{ x Y_1'(y) +  Y_2(y)}{1+ xY'(y)}+Y_1(y)$\\  \hline  \end{tabular}\end{center}
$\textrm{where $h, h_1$ are holomorphic functions of  the variable $z:= x+ i\cdot y$.}$
\end{Th}

{\bf Proof.}
It is well known (see, for example, \cite{Benenti3}),  that the condition $\{H, F\}=0$ is in this case equivalent to the following two conditions:
\begin{eqnarray} \label{cond1}
\{H_g, F_g\} & =&0 \, , \\
\label{cond2} 2 dU \circ G & =& dV \, , \end{eqnarray}
where $G$ is given by \eqref{matrix}. In tensor index notations, \eqref{cond2} is
\begin{equation} \label{keq}
2 G^i_j \frac{\partial U}{\partial x^i}=\frac{\partial V}{\partial x^j} \, .
\end{equation}
Indeed, condition $\{H, F\}=0$ is equivalent to the following equation:
$$\{H_g, F_g\}+ \{H_g, V\} - \{F_g, U\}  =0.$$
Since $\{H_g, F_g\}$ (respectively, $ \{H_g, V\} - \{F_g, U\}$) is a
third degree-polynomial in momenta
 (respectively, first degree), the latter equation is equivalent to:
 \begin{eqnarray} \label{cond1.}
\{H_g, F_g\} & =&0 \\
\{F_g, U\}   & =&\{H_g, V\} \, . \label{cond2.}  \end{eqnarray}
We see that \eqref{cond1.} coincides with \eqref{cond1} and \eqref{cond2.} is equivalent to
$$2 F^{ij} \frac{\partial U}{\partial x^i} =g^{ij} \frac{\partial V}{\partial x^i}  \, ,$$
which is equivalent to   \eqref{keq} and therefore to \eqref{cond2}.

Condition $ \eqref{cond1} $ tells us that the function
 $F_g$ is an integral quadratic in momenta
   for the geodesic flow of $g$. Clearly, $F_g$ is nontrivial. Indeed, if
   $F_g = \const_1\cdot  H_g$, then condition \eqref{cond2} reads $\const_1\circ dU   = dV  $ implying
   $V =  \const_1 \cdot U  + \const_2$. These in turn imply
   $F= \const_1\cdot  H+   \const_2$, which contradicts the assumptions.

   Thus, $F_g $ is a nontrivial integral of
   the geodesic flow of the metric $g$. By Theorem~\ref{main}, almost every point has a neighbourhood  with local coordinates $(x,y)$ such that
   $g$ and $F_g$ are as in the table. In order to prove Theorem~\ref{potential}, it is sufficient to show that, for every column of the table, the functions $U$ and $V$ are complete solutions of equation \eqref{cond2}.  Here we consider the three cases in detail. \weg{We will use that the condition \eqref{cond2} implies that the form $dU\circ G$ is closed.  }

   {\bf Liouville case. } Assume $g, F_g$ are as in the first column of the table. Then the form  $dU\circ G$ is
   $$-Y(y) \frac{\partial U}{\partial x}  dx - X(x) \frac{\partial U}{\partial y} dy $$
   and condition \eqref{cond2} reads
   \begin{equation}\label{li}  \left\{\begin{array}{ccc}
  \frac{ \partial Y(y) U}{\partial x}& =& -\tfrac{1}{2}\frac{\partial V}{\partial x} \, ,  \\
   \frac{ \partial X(x) U}{\partial y}& =&-\tfrac{1}{2} \frac{\partial V}{\partial y} \, .
   \end{array}\right. \end{equation}
  Differentiating the second equation w.r.t. $x$ and subtracting the derivative of the first equation w.r.t. $y$, we obtain
  $$0=  \frac{\partial }{\partial x}\left(X(x)  \frac{\partial U}{\partial y}\right)-  \frac{\partial }{\partial y}\left(Y(y) \frac{\partial U}{\partial x}\right)=  \frac{\partial^2 (X(x) -Y(y))U}{\partial x \partial y} $$
  implying
  $$U  = \tfrac{1}{2} \frac{\hat X(x) -\hat Y(y)}{X(x)- Y(y)} $$
  for certain functions $\hat X= \hat X(x) $ and $\hat Y = \hat Y(y)$.
Substituting $U$ in  \eqref{li}, we obtain
  $$
  V  = \frac{  X(x) \hat Y(y)-Y(y)  \hat X(x)}{X(x)-Y(y)}  \, . $$
     Thus, in the Liouville case, $U$ and $V$ are as in the table.

   {\bf Complex-Liouville case. }   In this case  $2 dU\circ G$  is equal to
   \begin{eqnarray*} &&
\left(\Re(h) \frac{\partial U}{\partial x} - \Im(h)
\frac{\partial U}{\partial y} \right)dx + \left(\Im(h) \frac{\partial U}{\partial x} + \Re(h)
\frac{\partial U}{\partial y} \right)dy\\ & =&  \left(\frac{\partial \Re(h)U}{\partial x} - \frac{\partial \Im(h)U}{\partial y}\right)dx + \left(\frac{\partial \Re(h)U}{\partial y} +  \frac{\partial \Im(h)U}{\partial x}\right)dy \end{eqnarray*}
and condition \eqref{cond2} is equivalent to the following system of PDE:
\begin{equation}\label{clc}  \left\{ \begin{array}{cc} \frac{\partial \Re(h)U}{\partial x} - \frac{\partial \Im(h)U}{\partial y} & = \frac{\partial V}{\partial x} \, ,\\
\frac{\partial \Re(h)U}{\partial y} +  \frac{\partial \Im(h)U}{\partial x}& = \frac{\partial V}{\partial y} \, .\end{array}
\right. \end{equation}
We see that these equation are precisely the Cauchy-Riemann condition for the function $h_1 : = \Re(h) U -V+ i\cdot \Im(h) U$. Thus,
$$ U= \frac{ \Im(h_1) }{\Im(h)}$$
and
$$V=  \Re(h) U - \Re(h_1)=  \Re(h)\frac{ \Im(h_1) }{\Im(h)}- \Re(h_1)\, . $$
We see that $U$ and $V $ are as in the table.

{\bf Jordan-block case.  }
 In this case the 1-form $2dU\circ G$ is
 $$-Y(y) \frac{\partial U}{\partial x} dx + \left((1+ x Y'(y))   \frac{\partial U}{\partial x} - Y(y) \frac{\partial U}{\partial y} \right)dy  $$
 and condition \eqref{cond2}  is equivalent to
the following system of PDE:
\begin{equation}\label{jlc}  \left\{ \begin{array}{cc} -Y (y)
\frac{\partial U}{\partial x}  & = \frac{\partial V}{\partial x}  \, ,\\
(1+ x Y'(y))   \frac{\partial U}{\partial x} - Y(y) \frac{\partial
U}{\partial y} & = \frac{\partial V}{\partial y}\, .
\end{array}\right.    \end{equation} The first equation in
\eqref{jlc} is equivalent to $ V= -Y(y) U + Y_1(y)$. Substituting
this in the second equation, we obtain
$$
(1+ x Y'(y))   \frac{\partial U}{\partial x} - Y(y) \frac{\partial U}{\partial y}= -\frac{\partial Y(y) U}{\partial y} + Y_1'(y)
$$
which implies
$$\frac{\partial (1+ xY'(y))U}{\partial x}= Y_1'(y) $$
and therefore $ (1+ xY'(y))U = x Y_1'(y) +  Y_2(y)$. Thus,
$$U= \frac{ x Y_1'(y) +  Y_2(y)}{1+ xY'(y)}$$
and
$$ V =  -Y \frac{ x Y_1'(y) +  Y_2(y)}{1+ xY'(y)}+Y_1(y)  \, . $$ \qed

\subsubsection{  Integration  by quadratures of    natural systems  admitting an integral quadratic in momenta }
\label{2.2.3}

Since the time of Jacobi it is known that (in  the 2-dimensional
Riemannian case) nontrivial   integrals  quadratic in momenta are
extremely helpful for the description of  dynamics of natural
systems: indeed, in this case
\begin{itemize} \item
the Hamilton equations,
which are a system of  four ODE on $T^*M^2$,
can be reduced  to a parameter-depending  system of   two  ODE
on $M^2$.\item    Moreover, it is possible  to construct a {\bf characteristic}  (= function constant on the solutions) of this system by means of the integration of certain functions of one variable only.  \end{itemize}
See \cite{3,Whi} for details.

    Classically,  the second  property is referred to as  {\bf  ``the system is integrable by quadratures".}
Both properties are useful for exact solutions, for numerical analysis and for a qualitative description of  (the solutions of)  the Hamilton equations.  We are going to show that these   nice   properties   persist
    in the  pseudo-Riemannian setting.

{\bf Liouville case.} There is virtually no  difference with respect to the Riemannian setting. Consider     $H= H_g + U $ and $F=F_g + V$ such that $g, F_g, U, V$ are as in the first column of the table from  Theorem~\ref{potential}.
Then, the first two Hamilton equations are
\begin{equation} \label{2h}\left\{
\begin{array}{ccccc}
\frac{d}{dt}  x& =& \frac{\partial H}{\partial p_x} &=& \frac{p_x}{X-Y} \, , \\
\frac{d}{dt}  y& =& \frac{\partial H}{\partial p_y} &=& -\frac{p_y}{X-Y} \, .
\end{array}\right.  \end{equation}

Since    the functions $F$ and $H$ are constant on the solutions  of the system, for every point $(x,y,p_x, p_y)$ of  the   solution we have

\begin{equation*} \left\{
\begin{array}{ccc}
 \tfrac{1}{2}\tfrac{p_x^2 -p_{y}^2}{X(x)-Y(y)} +
 \tfrac{1}{2}\tfrac{\hat X(x)  - \hat Y(y)}{X(x)-Y(y)}&=& H_0 \, , \\
 \tfrac{X(x)p_y^2 -Y(y)p_{x}^2}{X(x)-Y(y)} +
 \tfrac{\hat Y(y)X(x)  -\hat X(x) Y(y)}{X(x)-Y(y)}&=& F_0 \, .
 \end{array} \right. \end{equation*}
This is a linear system on $p_x^2, p_y^2$, solving it w.r.t. $p_x$ and $p_y$ we obtain
\begin{equation} \label{px}\left\{
\begin{array}{ccc}
p_x^2 &=& 2 H_0 X(x)   + F_0 - \hat X(x)\, , \\
p_y^2 &=& 2 H_0 Y(y)   + F_0 - \hat Y(y)\, .    \end{array}\right.
\end{equation}
Substituting these in (\ref{2h}), we obtain
\begin{equation}\label{s1} \left\{
\begin{array}{ccccc}
\frac{d}{dt}  x &=&  \varepsilon_1
\frac{ {\sqrt{ 2 H_0 X(x)   + F_0 - \hat X(x) }}}{X-Y} &:= &  v_1 \, , \\
\frac{d}{dt}  y & =&  \varepsilon_2
\frac{ {\sqrt {2 H_0 Y(y)   + F_0 - \hat Y(y) }}}{X-Y}  &:= &  v_2 \, .
\end{array}\right.
\end{equation}
We see that Hamilton equations can be reduced to a system of two ODE on $M^2$ depending on the parameters $H_0, F_0\in \mathbb{R} $ and $\varepsilon_i\in \{-1, +1\}$.

Clearly, a function $K(x,y) $ is a characteristic of the system \eqref{s1} if $dK$ vanishes on the vector field $v:= (v_1, v_2)$. Since the  form
$$B: =\frac{\varepsilon_1dx}{\sqrt {2 H_0 X(x)   + F_0 - \hat X(x) }} - \frac{\varepsilon_2dy}{\sqrt {2 H_0 Y(y)   + F_0 - \hat Y(y) }} $$ vanishes on $v$ and is closed, the function
$$K(p):=
\int_{p_0}^p  B =
\int_{x_0}^x \frac{d\xi}{\sqrt {2 H_0 X(\xi)   + F_0 - \hat X(\xi)}}  - \varepsilon_1\varepsilon_2
\int_{y_0}^y \frac{d\xi}{\sqrt {2 H_0 Y(\xi)   + F_0 - \hat Y(\xi)}}$$
is  a characteristic. We see that in order to find a characteristic, we only need to integrate two functions of one variable each,  i.e., the system is integrable by quadratures.

{\bf Complex-Liouville case.}  Consider     $H= H_g + U $ and $F=F_g + V$ such that $g, F_g, U, V$ are as in the second  column
of the table from  Theorem~\ref{potential}.
Then, the first two Hamilton equations are
\begin{equation} \label{2hc}\left\{
\begin{array}{ccccc}
\frac{d}{dt}  x& =&
\frac{\partial H}{\partial p_x} &=&
\frac{2p_y}{\Im(h)} \, , \\
\frac{d}{dt}  y& =&
\frac{\partial H}{\partial p_y} &=&
\frac{2p_x}{\Im(h)}\, .\end{array}\right.  \end{equation}

Since    the functions $F$ and $H$ are constant on the solutions  of the system,  for every point $(x,y,p_x, p_y)$ of  the solution we have

\begin{equation*} \left\{
\begin{array}{ccc}
2  \tfrac{p_xp_y}{\Im(h)} +  \frac{\Im(h_1)}{\Im(h)} &=& H_0 \, , \\
p_x^2 -p_y^2 +  \Re(h) \left(2\tfrac{p_xp_y}{\Im(h)} +  \frac{\Im(h_1)}{\Im(h)}\right) -  \Re(h_1)  &=& F_0 \, .\end{array} \right. \end{equation*}

 Subtracting the first equation  times $\Re(h)$  from  the second, we obtain
 \begin{equation*} \left\{
\begin{array}{ccc}
 2 {p_xp_y}   &=& H_0{\Im(h)}-  {\Im(h_1)}  \, ,
  \\
  p_x^2 -p_y^2  &=&   -\left(\Re(h) H_0 - \Re(h_1)\right) + F_0  \, .
  \end{array} \right. \end{equation*}
 From these, adding (respectively, substracting)  to (respectively, from)  the second equation  the first equation times $ i $,  we obtain
 \begin{equation*}\left\{
\begin{array}{ccccc}
\left( {p_x- i\cdot p_y}{}\right)^2   & = &  -  \left(H_0\Re(h)  -  \Re(h_1) -  {F_0}{}\right) - i\cdot  \left(H_0\Im(h) -  \Im(h_1)\right) &=&  -H_0 h+  h_1 + {F_0}{}  \, ,\\
 \left({p_x+ i\cdot p_y}{}\right)^2   & = & -  \left( H_0 \Re(h)-   \Re(h_1) - {F_0}{}\right) +  i\cdot   \left(H_0\Im(h)  -   \Im(h_1)\right) & =  &  - H_0 \bar h +  \bar h_1 + {F_0}{}  \, .
 \end{array} \right. \end{equation*}
 \begin{Rem} Since $\tfrac12 (p_x - i\cdot p_y)$  is the canonical momentum conjugate to $z = x + i\cdot y$,  these equations are the complex analog of \eqref{px}.
 \end{Rem}
 Then, $p_x= \varepsilon  \Re\left(\sqrt{ -H_0 h+  h_1 + {F_0}{}} \right) $ and $p_y= -\varepsilon   \Im\left(\sqrt{ -H_0 h+  h_1 + {F_0}{}} \right) $ (the choice of the branch of the square root is hidden in $\varepsilon$).
 Substituting these in \eqref{2hc}, we obtain
  \begin{equation} \label{s2}  \left\{
\begin{array}{ccccc}
\frac{d}{dt}  x &=&  \frac{ - 2\varepsilon \Im\left(\sqrt{ -H_0 h+  h_1 + {F_0}{}} \right)}{\Im(h)} & :=  & v_1  \, ,\\
\frac{d}{dt}  y & =&  \frac{ 2\varepsilon   \Re\left(\sqrt{ -H_0 h+  h_1 + {F_0}{}} \right)}{\Im(h)} & :=  & v_2  \, .\end{array}\right.
\end{equation}
We see that Hamilton equations can be reduced to a system of two ODE on $M^2$ depending on the parameters $H_0,  F_0\in \mathbb{R}, $ and $ \varepsilon \in\{-1, +1\}$.

Consider the 1-form
$$B :=
\frac{\Re\left(\sqrt{ -H_0 h+  h_1 + {F_0}{}} \right)}{|-H_0 h+  h_1 + {F_0}{}|}dx +  \frac{\Im\left(\sqrt{ -H_0 h+  h_1 + {F_0}{}} \right)}{|-H_0 h+  h_1 + {F_0}{}|}dy  \, .
$$
The Cauchy-Riemann conditions for the   holomorphic function $\sqrt{ -H_0 h+  h_1 + F_0} $
imply that the form is closed. Clearly, the  form vanishes on the vector field $v=(v_1, v_2)$. Then,
 the function
 $$K(p):=
 \int_{p_0}^{p} B =
 \int_{x_0}^x \frac{\Re\left(\sqrt{ -H_0 h+  h_1 + {F_0}{}} \right)}{|-H_0 h+  h_1 + {F_0}{}|}d\xi +
 \int_{y_0}^y \frac{\Im\left(\sqrt{ -H_0 h+  h_1 + {F_0}{}} \right)}{|-H_0 h+  h_1 + {F_0}{}|}d\xi $$
 is constant on
 the solutions of \eqref{s2}, i.e., is a characteristic of the system.
It is easy to check by  direct calculations  that in the complex coordinate $z$ the form  $B$ is
$$ 2 \Re\left( \frac{dz}{ \sqrt{-H_0 h+  h_1 + {F_0}{}}}\right) \, .$$
Thus, the function
$K$ equals to
$$ 2
\Re \left(\int_{z_0}^z  \frac{d\xi}{ \sqrt{-H_0 h(\xi)+  h_1(\xi) + {F_0}{}}}\right) \, , $$ i.e., the system is integrable by quadratures.

{\bf Jordan-block case.}  Consider     $H= H_g + U $ and $F=F_g + V$ such that $g, F_g, U, V$ are as in the third   column
of the table from  Theorem~\ref{potential}.
Then, the first two Hamilton equations are
\begin{equation} \label{3hc}\left\{
\begin{array}{ccccc}
\frac{d}{dt}  x& =& \frac{\partial H}{\partial p_x} &=& \frac{2p_y}{1 + x Y'(y)} \, ,\\ \frac{d}{dt}  y& =& \frac{\partial H}{\partial p_y} &=& \frac{2p_x}{1 + x Y'(y)} \, .
\end{array}\right.  \end{equation}

Since    the functions $F$ and $H$ are constant on the solutions  of the system,  for every point $(x,y,p_x, p_y)$ of  the solution we have

\begin{equation*} \left\{
\begin{array}{ccc}
2  \tfrac{p_xp_y}{1 + x Y'(y)} +  \frac{Y_2(y) + x Y_1'(y)}{1 + x Y'(y)} &=& H_0 \, ,\\  p_x^2 -  Y(y)  \left(2\tfrac{p_xp_y}{1 + x Y'(y)} +  \frac{Y_2(y) + x Y_1'(y)}{1 + x Y'(y)}\right) +  Y_1(y)  &=& F_0 \, .\end{array} \right. \end{equation*}

 Adding the first equation  times $Y(y)$  to   the second one,  we obtain
 \begin{equation*} \left\{
\begin{array}{ccc}
 {p_x^2}   &=& H_0Y(y) -  {Y_1(y)}+ F_0
  \\  2 p_x p_y   &=&   x\left(H_0Y'(y) -  {Y'_1(y)} \right) +     H_0   -Y_2(y)
  \end{array} \right.   \Longrightarrow
  \left\{ \begin{array}{ccc}
 p_x& = &\varepsilon   \sqrt{H_0Y(y) -  {Y_1(y)}+ F_0}  \, ,\\
 p_y& = &\tfrac{\varepsilon}{2}  \frac{  x\left(H_0Y'(y) -  {Y'_1(y)} \right) +     H_0   -Y_2(y)}{ \sqrt{H_0Y(y) -  {Y_1(y)}+ F_0  }} \, ,
 \end{array} \right.  \end{equation*}
where $\varepsilon\in \{-1, +1\}$.  Substituting these in \eqref{3hc}, we obtain
  \begin{equation} \label{s3}  \left\{
\begin{array}{ccccc}
\frac{d}{dt}  x &=&
\varepsilon \frac{  x\left(H_0Y'(y) -  {Y'_1(y)} \right) +     H_0   -Y_2(y)}
{ \left(1 + x Y'(y)\right)\sqrt{H_0Y(y) -  {Y_1(y)}+ F_0   }} & :=  & v_1 \, ,\\
\frac{d}{dt}  y & =&   \varepsilon\frac{ 2   \sqrt{H_0Y(y) -  {Y_1(y)}+ F_0  } }{1 + x Y'(y)} & :=  & v_2 \, .\end{array}\right.
\end{equation}
We see that Hamilton equations can be reduced to a system of two ODE on $M^2$ depending on the parameters $H_0,  F_0\in \mathbb{R}, $ and $ \varepsilon \in\{-1, +1\}$.

Consider the 1-form
\begin{eqnarray} \label{ar1} B &:=&
\frac{dx}{\sqrt{H_0Y(y) -  {Y_1(y)}+ F_0}}-
\frac12
\frac{x\left(H_0Y'(y) -  Y'_1(y)\right)- Y_2(y)+ H_0}
{\left(H_0Y(y) -Y_1(y)+ F_0\right)^{3/2}}dy  \\ &   =& d \left[
\frac{x}{\sqrt{H_0Y(y) -  {Y_1(y)}+ F_0}} \right] +
\frac12 \frac{ Y_2(y) - H_0}{\left(H_0Y(y) -  {Y_1(y)}+ F_0 \right)^{3/2}}dy \, .  \label{ar2} \end{eqnarray}

 By \eqref{ar2},  the  form   is closed. By   \eqref{ar1}, the form
 vanishes on the vector field $v=(v_1, v_2)$. Then,
 the function
 $$K(p):= \int_{p_0}^{p} B =
 \frac{x}{\sqrt{F_0 - Y_1(y) + H_0Y(y)}}\bigg\vert_{p_0}^{p} +  \frac12
 \int_{y_0}^{y}  \frac{ Y_2(\xi) - H_0}{\left(F_0 - Y_1(\xi) + H_0Y(\xi)\right)^{3/2}}d\xi  $$ is a characteristic of the system \eqref{s3},
  i.e. the system is integrable by quadratures.   \label{har}

\subsubsection{ Quantum integrability}\label{quant}

Let $g$ be a metric, and $(F^{ij})\in \Gamma(S_{2} M^2)$ be a  symmetric bilinear 2-form on $T^{*}M^2$. Consider the following two  linear partial differential
  operators $\Delta_g, \mathcal{F}_g: C^{\infty} \to C^{\infty}$:\\
  \begin{eqnarray*} \Delta_g&: =& -\sum_{i,j}\frac{1}{\sqrt{|det(g)|}}
    \frac{\partial }{\partial x_i} g^{ij} {\sqrt{|det(g)|}} \frac{\partial }{\partial x_j} \\
\mathcal{F}& :=& \sum_{i,j}\frac{1}{\sqrt{|det(g)|}}
    \frac{\partial }{\partial x_i} F^{ij} {\sqrt{|det(g)|}} \frac{\partial }{\partial x_j}
\end{eqnarray*}

\begin{Rem}  The first operator is the Beltrami-Laplace operator
 of the metric $g$; another way to write it down is
 $$
 \Delta_g = - \sum_{i,j}  g^{ij}\nabla_i \nabla_j ,
 $$
 where $\nabla$ is the Levi-C\`{\i}vita connection of $g$.  The second operator is a natural quantization of the function $\sum_{i,j} F^{ij} p_ip_j$ and another  way to write it down is
 $$
 \mathcal{F}_g = \sum_{i,j}  \nabla_i F^{ij}\nabla_j .
 $$
 In particular, both operators do not depend on the choice of the coordinate system.
 \end{Rem}

 \begin{Rem} The symbols of $\Delta_g$ and of $\mathcal{F}_g$ are $-2H:=-2\,  \sum_{i,j} g^{ij} p_ip_j $ and $\sum_{i,j} F^{ij} p_ip_j$, respectively.  \end{Rem}

\begin{Th} \label{quantum}
Let  $F= \sum_{i,j}F^{ij}  p_ip_j+V(x,y)$ be
 a  quadratic  integral of the natural Hamiltonian system $\frac12 \sum_{i,j}g^{ij}  p_ip_j+U(x,y)$ on $T^*M^2$. Then, the operators
$$ \mathcal{H} := \Delta_g - 2 U $$
 and
 $$ \mathcal{F} := \mathcal{F}_g  + V $$
 commute:  $\mathcal{H} \circ \mathcal{F} = \mathcal{F} \circ\mathcal{H}$.
 \end{Th}

 \begin{Rem}  The Riemannian analog of Theorem \ref{quantum} follows from
 \cite{benenti,era1,Quantum,quantum}.
 \end{Rem}

{ \bf Proof of Theorem \ref{quantum}. }
 It is sufficient to check the statement at almost every point, i.e.,
  for the metrics and the integrals from Theorem~\ref{potential}. Direct calculations shows that in this case   the operators $\Delta_g $ and $\mathcal{F}_g$ are as in the following table:

 \begin{center} \begin{tabular}{|c||c|c|c|}\hline & \textit{Liouville case} & \textit{Complex-Liouville case} & \textit{Jordan-block case}\\ \hline \hline
$\Delta_g$ & $ \frac{- 1}{X \left( x \right) -Y\left( y \right) }\left(\frac {\partial ^{2}}{\partial x^{2}} -{
\frac{\partial^2}{\partial y^2}
 }\right)
$ &  $\frac{-4}{{\Im(h)}  }  {{\frac {\partial ^{2}}{\partial x{{}}\partial y{{}}}}
 }
$ & ${\frac{-4}{1+x_{{1}}{ {}{{{}}}}Y'
 \left( y \right) }{\frac {\partial ^{2}}{\partial x{{}}\partial y{{}}}}
 }
 $
\\  \hline  $ \mathcal{F}_g $  &   $ \frac{1}{X \left( x{
{}} \right)     -Y \left( y{{}} \right) }\left( X \left( x \right)\frac{\partial^{2}}{\partial y^2}- Y \left( y \right) \frac{\partial ^{2}} {\partial {x^{2}}}
   \right)
$&
 $\frac{\partial^2 }{\partial x^2} - \frac{\partial^2 }{\partial y^2} + 2\frac{\Re(h)}{\Im(h)} \frac{\partial^2 }{ \partial x \partial y} $
&  $  \frac{\partial^2}{\partial x^2}  -2  \frac{Y(y)}{1+ {x}{} Y'(y) }\frac{\partial^2 }{ \partial x \partial y}$\\ \hline \end{tabular}\end{center}
where $h$ is a holomorphic function of $z = x + i\cdot y $.

 To prove that  $\mathcal H=\Delta_g -2U$ and $\mathcal F=\mathcal F_g + V$ commute, we first observe
 that in the Liouville and Jordan-block cases:
 $$\mathcal F_g + V=\frac{\partial^2}{\partial x^2}+ f\cdot (\Delta_g-2U) + f_1,$$
 where $f=X(x)$,  $f_1=\hat X(x)$  for the Liouville case, and
 $f=\frac{Y(y)}{2}$ and $f_1=Y_1(y)$ for the Jordan block case.

 Similarly, in the complex Liouville case, we have
 $$\mathcal F_g + V=\frac{\partial^2}{\partial x^2}-\frac{\partial^2}{\partial y^2}+ f\cdot (\Delta_g-2U) + f_1$$
 where $f=-\frac{\Re (h)}{2}$, $f_1=\frac{\Re (h_1)}{2}$

The Laplace-Beltrami operator $\Delta_g$ in all the cases is of the
form $\Delta_g=\lambda^{-1}  \Delta_{g_0}$, where $\Delta_{g_0}$ is
the Laplace-Beltrami operator of the flat metric $g_0$  (more
specifically, $g_0$ is $dx^2 - dy^2$ in the Liouville case, and
$2dxdy$ in the complex Liouville and Jordan-block cases). Using the
fact that $\Delta_{g_0}$ commutes with $\frac{\partial}{\partial
x}$, it is straightforward to verify the following commutator
formula:
$$
[ \Delta_g -2U, \frac{\partial^2}{\partial x^2} ]=
  \left(  \frac{\lambda _{xx} }{\lambda}
 +
 2 \frac{\lambda_x}{\lambda}  \frac{\partial }{\partial x} \right)  \circ (\Delta_g-2U) +2
\frac{(\lambda U)_{xx}}{\lambda}+ 4 \frac{(\lambda U)_x}{\lambda}
\frac{\partial }{\partial x}
$$
(here we use standard notation for the commutator of two linear
operators $[\mathcal A,\mathcal B]=\mathcal A\circ \mathcal B -
\mathcal B\circ \mathcal A$).

The two following formulas are standard:
$$
[\Delta_g -2 U ,  f \cdot (\Delta_g-2U)] = \left( \Delta_g f - 2\,
{\text{grad}_g f} \right) \circ (\Delta_g-2U)
$$
and
$$
[\Delta_g-2U, f_1]= \Delta_g f_1 - 2 \, {\text{grad}_g f_1},
$$
where the vector field ${\text{grad}_g f}$ is viewed as a first
order differential operator, i.e., $\text{grad}_g f=g^{ij}
\frac{\partial f}{\partial x^i} \frac{\partial}{\partial x^j}$.

Thus,  in the Liouville and Jordan-block cases,  we have:
$$
\begin{aligned}{}
[\mathcal H, \mathcal F]=[\Delta_g -2U, \mathcal F_g + V]&=\left(
\frac{\lambda _{xx}}{\lambda} +
 2 \frac{\lambda_x}{\lambda}  \frac{\partial }{\partial x} + \Delta_g f - 2\, {\text{grad}_g f}  \right)
 \circ (\Delta_g-2U) +
 \\
 &+ 2
\frac{(\lambda U)_{xx}}{\lambda}+ 4 \frac{(\lambda U)_x}{\lambda}
\frac{\partial }{\partial x} + \Delta_g f_1 - 2\, { \text{grad}_g
f_1}
\end{aligned}
$$
Hence, the commutativity condition $[\mathcal H, \mathcal F]=
\mathcal H\circ\mathcal F -\mathcal H\circ\mathcal F =0$ splits into
four simple equations (here we use the fact that $\Delta_g =
\lambda^{-1} \Delta_{g_0}$ and $\text{grad}_g =
\lambda^{-1}\text{grad}_{g_0})$:
$$
\begin{array}{ll}
\text{(i)} & {\lambda_x} \frac{\partial }{\partial x}-{\text{grad}_{g_0} f} =0\\
\text{(ii)} & {\lambda _{xx} }  + \Delta_{g_0} f =0\\
\text{(iii)} & 2{(\lambda U)_x}
\frac{\partial }{\partial x} - { \text{grad}_{g_0}f_1}=0\\
\text{(iv)} & 2{(\lambda U)_{xx}}+ \Delta_{g_0} f_1 =0
\end{array}
$$
Each of these equations has natural meaning. Indeed, (i) and (ii)
mean that the operators $\Delta_g$ and ${\mathcal F}_g$  commute
(without potentials), (iii) and  (iv) give the ``new" commutativity
conditions involving the potentials. The first and third equations
are equivalent to the commutativity of classical integrals, whereas
the second and the fourth keep additional ``quantum'' information.
It is interesting to notice that the quantum conditions (ii) and
(iv) can be obtained from the classical ones (i) and (iii) by
``differentiating" so that in our particular case the quantum
integrability in dimension 2 turns out to be a corollary of the
classical one:
$$
\begin{array}{l}
{\lambda _{xx} }  + \Delta_{g_0} f = \mbox{div} ({\lambda_x}
\frac{\partial }{\partial x}-{\text{grad}_{g_0} f} )\\
2{(\lambda U)_{xx}}+ \Delta_{g_0} f_1 = \mbox{div}(2{(\lambda U)_x}
\frac{\partial }{\partial x} - { \text{grad}_{g_0}f_1} ) \end{array}
$$

However, each of the above four conditions can be verified directly.
Taking into account the following explicit formulas:
$$
\begin{array}{llll}
\text{Liouville case}: & &  \\
\Delta_{g_0}= -\frac
{\partial}{\partial x^2} + \frac {\partial}{\partial y^2}, &
\text{grad}_{g_0} f = f_x \frac {\partial}{\partial x} - f_y \frac
{\partial}{\partial
y}, & \lambda=X(x)-Y(y) \\
\text{Jordan-block case}: & & \\
\Delta_{g_0}= -2\frac {\partial^2}{\partial x\partial y}, &
\text{grad}_{g_0} f = f_y \frac {\partial}{\partial x} + f_x \frac
{\partial}{\partial y}, & \lambda=1/2 (1+ xY'(y))
\end{array}
$$
we see that equations(i)--(iv) become:
$$
\begin{array}{l}
\text{Liouville case:}\\
\qquad ({\lambda_x} - f_x) \frac {\partial}{\partial x} + f_y \frac {\partial}{\partial y} =0\\
\qquad {\lambda _{xx} }  -  f_{xx} + f_{yy} =0\\
\qquad (2 {(\lambda U)_x}  -  (f_1)_x) \frac {\partial}{\partial x} + (f_1)_y \frac {\partial}{\partial y}=0\\
\qquad {2(\lambda U)_{xx}}- (f_1)_{xx} + (f_1)_{yy} =0
\end{array}
\qquad
\begin{array}{l}
\text{Jordan-block case:}\\
\qquad ( {\lambda_x} - f_y)\frac{\partial }{\partial x}-f_x \frac
{\partial}{\partial y}  =0\\
\qquad {\lambda _{xx} }  - 2f_{xy} =0\\
\qquad (2 {(\lambda U)_x} - (f_1)_y) \frac{\partial }{\partial x} -
(f_1)_x \frac {\partial}{\partial y} =0\\
\qquad {2(\lambda U)_{xx}}- 2(f_1)_{xy} =0
\end{array}
$$
and obviously hold for $\lambda$, $f$ and $f_1$ indicated above.

The complex Liouville case is absolutely similar, the only difference is the additional term
$\frac{\partial^2}{\partial y^2}$, which  leads to the following system of relations:
$$
\begin{array}{l}
\text{Complex Liouville case:}\\
\qquad ( {\lambda_x} - f_y)\frac{\partial }{\partial x}+ (-\lambda_y-f_x) \frac {\partial}{\partial y}  =0\\
\qquad {\lambda _{xx} } - \lambda_{yy} - 2 f_{xy} =0\\
\qquad (2 {(\lambda U)_x} - (f_1)_y) \frac{\partial }{\partial x} + (-2(\lambda U)_y -(f_1)_x) \frac {\partial}{\partial y}  =0\\
\qquad {2(\lambda U)_{xx}}-2{(\lambda U)_{yy}}- 2(f_1)_{xy} =0
\end{array}
$$
each of which obviously holds for $f=-\frac{\Re (h)}{2}$, $f_1=-\Re
(h_1)$, $\lambda=\frac{\Im (h)}{2}$, $2\lambda U=\Im (h_2)$. \qed

\section{Proof of Theorem \ref{main}}
\subsection{Admissible coordinate systems and
Birkhoff-Kolokoltsov forms} \label{admissible}

Let $g$ be a pseudo-Riemannian metric on $M^2$  of signature $(+, -)$.
Consider (and fix)  two vector fields $V_1, V_2$ on $M^2$  such that
\begin{itemize}
 \item  $g(V_1, V_1) =g(V_2, V_2)=0$ and
 \item $g(V_1, V_2)>0$.
 \end{itemize}
Such vector fields always exist locally, (and since our result is local, this is sufficient for our proof). For possible further use, let us note that such vector fields always exist   on a finite (at most, 4-sheet-) cover of $M^2$.

We will say that a local  coordinate system $(x,y)$
 is {\bf admissible}, if  the vector fields  $\frac{\partial }{\partial x}$ and  $\frac{\partial }{\partial y} $ are proportional to $V_1, V_2$ with positive coefficient of proportionality:  $$\frac{\partial }{\partial x}=  \lambda_1(x,y) V_1(x,y), \  \  \  \frac{\partial }{\partial y}=  \lambda_2(x,y) V_2(x,y), \  \  \ \textrm{where $\lambda_i>0$}.$$
 Obviously,
\begin{itemize}
\item  admissible coordinates exist in a sufficiently small neighbourhood of every point, \item the metric $g$  in  admissible coordinates has the form
 \begin{equation}\label{metric}
 ds^2 =f(x,y)dxdy , \  \  \ \textrm{where $f>0$},  \end{equation}
     \item two admissible  coordinate systems  in
      one neighbourhood  are connected by \begin{equation} \label{coordinatechange} \begin{pmatrix} x_{new}\\
 y_{new}\end{pmatrix}
 = \begin{pmatrix} x_{new}(x_{old}) \\
 y_{new}(y_{old})\end{pmatrix} , \  \ \textrm{where  $\frac{dx_{ new}}{dx_{old}}>0$, $\frac{dy_{ new}}{dy_{old}}>0$}. \end{equation}
\end{itemize}

\begin{Lemma}  \label{BK}
Let $(x,y)$ be an admissible coordinate system for $g$.
Let  $F$ given by  \eqref{integral} be  an  integral for  $g$.
Then,
$$
B_1:= \frac{1}{\sqrt{|a(x,y)|}}dx, \;\; \left({\rm respectively },
B_2:= \frac{1}{\sqrt{|c(x,y)|}}dy \right)
$$
is a    1-form, which is defined at points such that $a\ne 0$ (respectively, $c\ne 0$).  Moreover, the coefficient
 $a$ (respectively, $c$) depends only on $x$ (respectively, $y$), which in particular implies that the forms $B_1$, $B_2$ are  closed.   \end{Lemma}

\begin{Rem}  The forms $B_1, B_2$ are not the direct analog of the ``Birkhoff" 2-form introduced by
Kolokoltsov in \cite{Kol}. In a certain sense, they are a real
analog of the  square root of the Birkhoff  form.
\end{Rem}

{\bf Proof of Lemma~\ref{BK}.} The first part of the statement, namely that $$
\frac{1}{\sqrt{|a(x,y)|}}dx, \;\; \left({\rm respectively },
\frac{1}{\sqrt{|c(x,y)|}}dy \right)
$$
transforms as a $1$-form under admissible coordinate changes is evident:  indeed, after the coordinate change
\eqref{coordinatechange}, the  momenta transform as follows:
 $p_{x_{old}}= p_{x_{new}}\frac{d{x_{new}}}{d{x_{old}}}$, $p_{x_{old}}= p_{x_{new}}\frac{d{x_{new}}}{d{x_{old}}}$. Then, the integral $F$ in the new coordinates has
  the form
  $$ \underbrace{\left(\frac{d{x_{new}}}{d{x_{old}}}\right)^2{a}}_{a_{new} } {p_{x_{new}}^2}  + \underbrace{\frac{d{x_{new}}}{d{x_{old}}}\frac{d{y_{new}}} {d{y_{old}}}{b}}_{b_{new}}  {p_{x_{new}}} {p_{y_{new}}} + \underbrace{\left(\frac{d{y_{new}}}{d_{y_{old}}} \right)^2{c}}_{c_{new}} {p_{y_{new}}^2}.$$
  Then, the formal expression $\frac{1}{\sqrt{|a|}}dx_{old}
  $ ({\rm respectively}, $ \frac{1}{\sqrt{|c|}}dy_{old}$)   transforms into
  $$
  \frac{1}{\sqrt{|a|}} \frac{d{x_{old}}}{d{x_{new}}} dx_{new} \ \ \ \ \  \left(\textrm{respectively,  $ \frac{1}{\sqrt{|c|}}\frac{d{y_{old}}}{d{y_{new}}}dy_{new}$}\right),  $$
  which is precisely the transformation law of  1-forms.

Let us prove that  the coefficient
 $a$ (respectively, $c$) depends only on $x$ (respectively, $y$), which in particular implies that the forms $B_1$, $B_2$ are  closed.
 If $g$ is given by \eqref{metric}, its Hamiltonian is
 $$H=\frac{2p_xp_y}{f} \, ,$$
 and the condition $\{H, F\}=0$ reads \\
 \begin{eqnarray*}
0&=& \left\{\frac{2p_xp_y}{f}, ap_x^2+ bp_xp_y+ cp_y^2\right\}  \\
 &=& \frac{2}{f^2}\left(p_x^3(fa_y) + p_x^2 p_y (fa_x + fb_y + 2 f_x a + f_y b)+ p_yp_x^2 (fb_x + fc_y+ f_x b + 2 f_y c)+ p_y^3 (c_xf)\right) \, ,
 \end{eqnarray*}
 i.e., is equivalent to the following system of PDE:
 \begin{equation}\label{sys}
 \left\{\begin{array}{rcc} a_y&=&0 \, ,\\
 fa_x + fb_y + 2 f_x a + f_y b&=&0\, ,\\
 fb_x + fc_y+ f_x b + 2 f_yc &=&0\, ,\\
 c_x&=&0 \, .\end{array}
 \right.\end{equation}

 Thus, $a=a(x)$, $c=c(y)$, which is equivalent to state that
  $B_1:= \frac{1}{\sqrt{|a|}} dx$ and $B_2:= \frac{1}{\sqrt{|c|}}dy$ are closed forms (assuming $a\ne 0$ and $c\ne 0$). \qed

\begin{Rem} \label{rem3} For further use let us formulate one more consequence of equations \eqref{sys}: if $a\equiv c \equiv 0$ in a neighbourhood of  a point, then $bf = \const$, implying     $F\equiv  \const \cdot H$ in the neighbourhood.
\end{Rem}

\vspace{1ex}
Assume $a\ne 0$  (respectively, $c\ne 0$)
 at a point $p_0$. For every $p_1$
in a small neighbourhood $U$ of $p_0$ consider
 \begin{equation} \label{normal}
 x_{new} :=\int\limits_{\begin{array}{c}
\gamma:[0,1]\to U \\  \end{array}}
 B_1, \ \ \left(\textrm{respectively, $ y_{new} :=\int\limits_{\begin{array}{c}
\gamma:[0,1]\to U \\  \end{array}}
B_2 $ }\right) \, ,\end{equation}
with $\gamma(0)=p_0, \gamma(1)= p_1$.

Locally, in the admissible coordinates,
 the functions $x_{new}$ and $y_{new}$ are given by
 \begin{equation}\label{loc:normal}  x_{new}(x{})=\int_{x_0}^{x_{}} \frac{1}{\sqrt{|a(t)|} }\, dt,  \ \ \  \   y_{new}(y)=\int_{y_0}^{y_{}} \frac{1}{\sqrt{|c(t)|} }\, dt \, .
\end{equation}

The coordinates $(x_{new},y_{old})$, $\bigl((x_{old}, y_{new})$,
$(x_{new}, y_{new})$, respectively$\bigr)$    are admissible.  In
these coordinates the forms $B_1, B_2$   are given by
 $dx_{new}$, $dy_{new}$ implying that
$a=c=\pm 1$ (more precisely: $a_{new}= \textrm{sign}(a_{old})$, $c_{new}= \textrm{sign}(c_{old})$).

\subsection{Proof of Theorem~\ref{main}}
We assume that $g$ on  $M^2$ of  signature (+,--)
  admits  a nontrivial quadratic integral $F$ given by \eqref{integral}. Consider the $(1,1)$-tensor $G$ given by \eqref{matrix}. In a neighbourhood of almost every point, the Jordan normal form of this $(1,1)$-tensor is one of the following:

 {Case 1} $\begin{pmatrix} \lambda & 0\\ 0 & \mu  \end{pmatrix} $, where $\lambda, \mu\in \mathbb{R}$.

 {Case 2}   $\begin{pmatrix} \lambda+ i \mu  & 0\\ 0 & \lambda- i \mu   \end{pmatrix} $, where $\lambda, \mu\in \mathbb{R}$.

  {Case 3} $\begin{pmatrix} \lambda  & 1\\ 0 & \lambda   \end{pmatrix} $, where $\lambda \in \mathbb{R}$.

 Moreover,  in view of Remark~\ref{rem3}, there exists a neighbourhood of almost every point such that $\lambda \ne \mu$ in case 1 and $\mu \ne 0 $ in case 2.
  In the admissible coordinates, up to multiplication of $F $ by $-1$,
  case 1 is equivalent to the condition
    $ac>0$, case 2 is equivalent to the condition   $ac<0$ and, finally, case 3 is equivalent to the   condition   $ac=0$.

  We now consider all three cases.

  \subsubsection{Case 1: $ac>0$. }
 Without loss of generality we assume $a>0$, $c>0$.
 Consider the    coordinates \eqref{normal}.
 In these coordinates  $a=1$, $c=1$ and equations \eqref{sys} have the following simple form.

 \begin{equation}\label{sys2}
 \left\{\begin{array}{rcc}  (fb)_y+ 2 f_x  &=&0 \,,\\
 (fb)_x  + 2 f_y&=&0 \,. \end{array}
 \right.\end{equation}
 This system can be solved. Indeed, it is equivalent to
 \begin{equation}\label{sys3}
 \left\{\begin{array}{rcc}  (fb+ 2 f)_x  + (fb  + 2 f)_y&=&0 \,,\\
   (fb  -2 f)_x  -(fb- 2 f)_y&=&0 \,, \end{array}
 \right.\end{equation}
 which after the (non-admissible) change of coordinates $x_{new} = x+y$, $y_{new}= x-y$, has the form
 \begin{equation}\label{sys4}
 \left\{\begin{array}{rcc}  (fb+ 2 f)_x&=&0 \,,\\
   (fb  -2 f)_y&=&0 \,, \end{array}
 \right.\end{equation}
 implying $fb+ 2f = Y(y)$, $fb-2f=X(x)$. Thus,
 $$
 f= \frac{Y(y)-X(x)}{4} \, , \;\; b=2 \frac{X(x)+Y(y)}{Y(y)-X(x)}\, .
 $$

 Finally, in the new coordinates, the metric and the integral have (up to a  possible multiplication by a constant) the form

 \begin{equation}\label{answer:case1}
 (X-Y)(dx^2 -dy^2) \, , \end{equation}
 \and
 \begin{equation}
 \frac12 \left(p_x^2 -  \tfrac{X(x)+Y(y)}{X(x)-Y(y)}(p_x^2-p_y^2) + p_y^2\right)=
 \frac{p_y^2 X(x) - p_x^2Y(y)}{X(x)-Y(y)}  \, .
 \end{equation}

  \subsubsection{ Case 2:
  $ac<0$. }

  Without loss of generality we can assume  $a>0$, $c<0$.
 Consider the normal coordinates \eqref{normal}.
 In these coordinates  $a=1$, $c=-1$ and equations \eqref{sys} have the following simple form.

 \begin{equation}\label{sys:case2}
 \left\{\begin{array}{rcc}  (fb)_y+ 2 f_x  &=&0 \,,\\
 (fb)_x  -2 f_y&=&0 \,. \end{array}
 \right.\end{equation}
 We see that these equations are the Cauchy-Riemann  conditions for the complex-valued
 function $fb+ 2i f$. Thus, for an appropriate holomorphic
 function $h= h(x+ iy)$ we have $fb=\Re(h)$, $2f =\Im(h)$.

 Finally, in a certain coordinate system, the metric and the integral are
 (up to possible multiplication by constants)
  \begin{equation}\label{answer:case2}
 \Im(h)dxdy \ \ \ \textrm{and} \ \ \ p_x^2 - p_y^2   + 2\frac{\Re(h)}{\Im(h)}p_xp_y
 \end{equation}

 \subsubsection{Case 3: $ac=0$.  }

 Without loss of generality we can assume  $a>0$, $c=0$.
 Consider  admissible
 coordinates $x, y$, such that  $x$ is the
 normal coordinate from  \eqref{normal}.
 In these coordinates  $a=1$, $c=0$, and the equations \eqref{sys} have the following simple form.
 \begin{equation}\label{sys:case3}
 \left\{\begin{array}{rcc}  (fb)_y+ 2 f_x  &=&0 \, ,\\
 (fb)_x  &=&0 \, .\end{array}
 \right.\end{equation}
 This system can be solved. Indeed, the second equation implies $fb= -Y(y)$. Substituting this in the first equation  we obtain
 $Y'= 2f_x$ implying
 $$f= \frac{x}{2} Y'(y)+ \widehat{ Y}(y) \textrm{ \ \ and}  \ \ \ b= - \frac{Y(y)}{\frac{x}{2} Y'(y)+ \widehat{ Y}(y)}\, .
 $$
 Finally, the metric and the integral are
  \begin{equation}\label{answer:case3}
 \left( \widehat{ Y}(y)+\frac{x}{2} Y'(y)\right)dxdy \ \ \ \textrm{and} \ \ \ p_x^2 - \frac{Y(y)}{\widehat{ Y}(y)+\frac{x}{2} Y'(y) }p_xp_y \, .
 \end{equation}

 Moreover, by the change $y_{new}=\beta(y_{old})$, equations \eqref{answer:case3} will be simply transformed to:
  \begin{equation}\label{answer:case3transformed}
 \left( \widehat{ Y}(y)\beta' +\frac{x}{2} Y'(y)\right)dxdy \ \ \ \textrm{and} \ \ \ p_x^2 -  \frac{Y(y)}{\widehat{ Y}(y)\beta'+\frac{x}{2} Y'(y) }p_xp_y  \, .
 \end{equation}
 Thus, by putting $\beta(y) = \int_{y_0}^y \frac{1}{\widehat{ Y}(t)} dt$,  we can make the metric and the integral to be
 $$  \left( 1+\frac{x}{2} Y'(y)\right)dxdy \ \ \ \textrm{and} \ \ \ p_x^2 -   \frac{Y(y)}{1+\frac{x}{2} Y'(y) }p_xp_y  \, .
 $$
 Moreover, after the coordinate change $x_{new} =  \tfrac{x_{old}}{2}$ and multiplication of the metric by $\tfrac{1}{2}$, the metric and the integral have  the form  from Theorem~\ref{main}
  \label{case3}
  \begin{equation}\label{answer:case3final}  \left( 1+ {x} Y'(y)\right)dxdy \ \ \ \textrm{and} \ \ \ p_x^2 -2  \frac{Y(y)}{1+ {x}{} Y'(y) }p_xp_y  \, .
 \end{equation}

  Theorem~\ref{main} is proved.

 \begin{Rem} \label{14} 
  Let us note  that if $dY\ne 0$, then we can take $Y$ as the coordinate $y$.  Then, the metric and the integral  \eqref{answer:case3} will have the form (see also Remark \ref{r4})   \begin{equation}\label{case3:special}  \left( \tilde Y(y) -\frac{x}{2} \right)dxdy \ \ \ \textrm{and} \ \ \ p_x^2 +  \frac{y}{\tilde Y(y)-\frac{x}{2}  }p_xp_y  \, .
 \end{equation}
  \end{Rem}

\section{Conclusions}
We have discussed integrable geodesic flows of pseudo-Riemannian metrics on 2-dimensional manifolds constructing (Theorem \ref{main}) local normal forms of
such metrics. The normal forms are of three types: Liouville (the analogous of the Riemannian case), Complex-Liouville and Jordan-block. We have shown that these metrics, in analogy with the Riemannian case, admit geodesically equivalent metrics, can be used to construct a large family of natural  systems admitting integrals quadratic in momenta, that these natural systems are integrable by quadratures and that the integrability of such systems can be generalized to the quantum setting. A natural further step in this field would be to understand what is the structure of the quadratic integral in the case in which the manifold is closed (the Riemannian case is done in \cite{1,BMF,Kio,Kol}).

\section*{Acknowledgement}  
The first author thanks the Russian Foundation for Basic Research (grants 05-01-00978) for partial financial support.
The second  author thanks
Deutsche Forschungsgemeinschaft
(Priority Program 1154 --- Global Differential Geometry) for partial financial support and Loughborough and Cambridge Universities, and also  Universit\`a di Roma ``Tor Vergata" for their  hospitality.
The third  author acknowledges the financial support from INFN, Sezione di Roma II.

 \end{document}